\documentclass{article}

\usepackage{microtype}
\usepackage{graphicx}
\usepackage{booktabs} 

\usepackage{hyperref}
\usepackage{xurl}

\usepackage[accepted, preprint]{icml2026}

\usepackage{amsmath}
\usepackage{amssymb}
\usepackage{printlen}
\usepackage{url}

\usepackage{subcaption}
\usepackage[ruled,vlined,linesnumbered,algo2e]{algorithm2e}
\usepackage{lipsum}

\icmltitlerunning{Efficient Safety Benchmarking via Item Response Theory}

\begin{document}

\twocolumn[
    \icmltitle{Efficient Safety Benchmarking via Item Response Theory}
    \icmlsetsymbol{equal}{\textbf{*}}
    \icmlsetsymbol{supervision}{\textbf{\ensuremath{\dagger}}}

    \begin{icmlauthorlist}
    \icmlauthor{Fabio Spagliardi}{equal,yyy}
    \icmlauthor{Mírian Silva}{equal,yyy}
    \icmlauthor{Ayan Datta}{equal,yyy}
    \icmlauthor{Aiden Zhou}{yyy,sch}
    \icmlauthor{Vamshi Bonagiri}{supervision,yyy,mbzuai}
    \icmlauthor{Diogo Cruz}{supervision,yyy}
    \end{icmlauthorlist}
    
    \icmlaffiliation{yyy}{Supervised Program for Alignment Research (SPAR)}
    \icmlaffiliation{mbzuai}{Mohamed bin Zayed University of Artificial Intelligence (MBZUAI)}
    \icmlaffiliation{sch}{Yale University}
    
    \icmlcorrespondingauthor{Fabio Spagliardi}{spagliardi.fabio@gmail.com}
    \icmlcorrespondingauthor{Mírian Silva}{mirianfrsilva@gmail.com}
    \icmlkeywords{AI safety}
    \vskip 0.3in
]
\printAffiliationsAndNotice{
\icmlEqualContribution
\textsuperscript{\ensuremath{\dagger}}Equal supervision.
}


\begin{abstract}
Safety benchmarks for language models are typically evaluated using static paradigms that treat all items as equally informative for all models, an assumption that is particularly problematic for adversarial, highly heterogeneous safety items. Applied in full to modern benchmark suites, current evaluation procedures would require on the order of $10^5$ responses, most of which provide little ranking signal. We analyze six widely used safety benchmarks and make three contributions toward more efficient safety evaluation. First, we show that Item Response Theory (IRT) recovers interpretable structure on safety benchmarks, with ability estimates resolving differences among models that cluster at the ceiling of raw safety metrics. Second, we show that adaptive item selection, which dynamically chooses informative items for each model based on its responses, approximates full-benchmark rankings (Spearman's $\rho >$ 0.90), reducing evaluation cost by at least 80\% on every benchmark where this threshold is attainable, and by up to 99.9\%, achieved on AIR-Bench 2024. Third, we introduce a practical procedure for extracting a fixed, informative subset of items reusable across models, a static alternative to adaptive selection with savings of 80--99.8\% across benchmarks. Together, these results establish that psychometric methods enable benchmark-aware reductions in evaluation costs across the safety evaluation pipeline.
\end{abstract}

\section{Introduction}
\label{sec:intro}


\begin{figure*}[ht]
    \centering
    \includegraphics[width=0.8\linewidth]{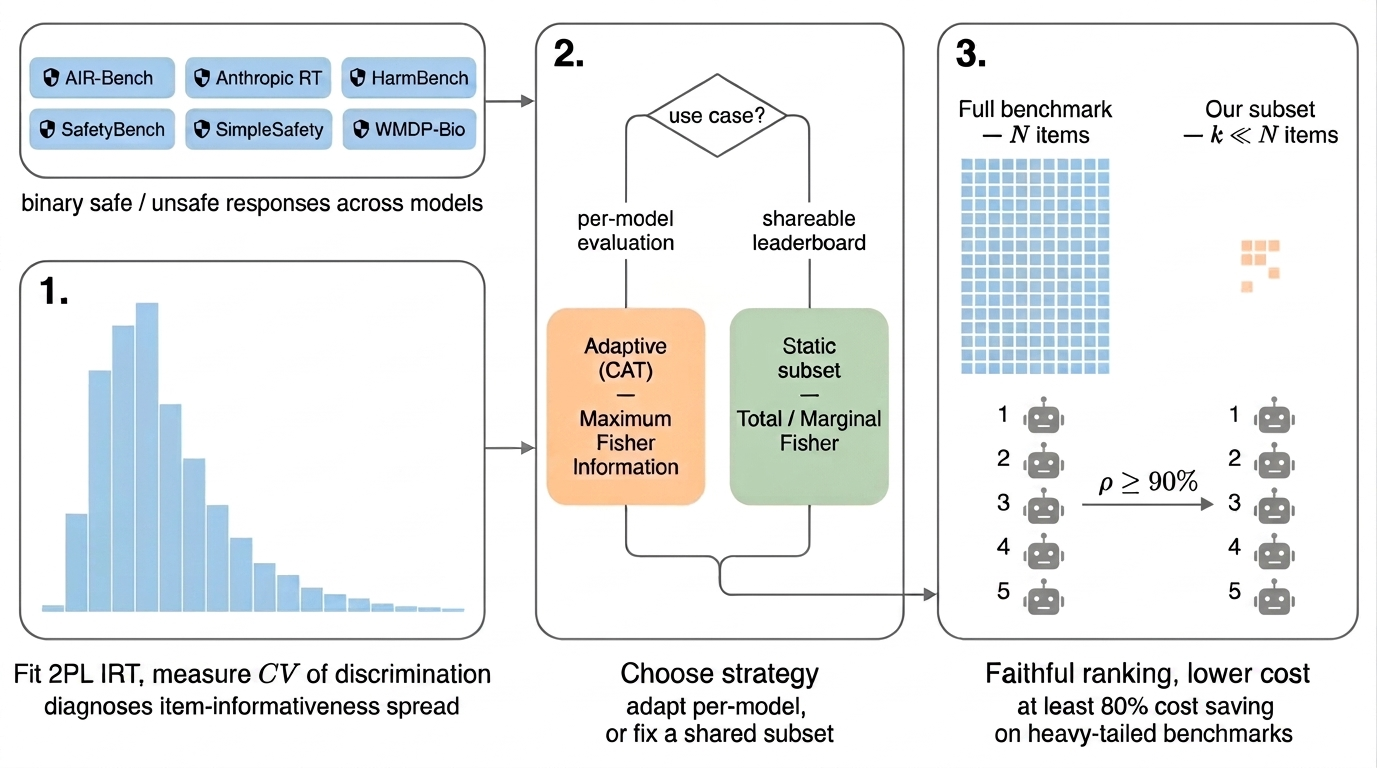}
    \caption{We apply Item Response Theory (IRT) to safety benchmarks, showing that it reveals interpretable structure in benchmark data. Building on this, we develop two item selection procedures: an adaptive approach that selects items tailored to each model to maximize information gain, and a static approach that defines an optimal benchmark subset applicable to all models without further modification. Both approaches accurately reproduce model rankings.}
    \label{fig:recipe__}
    \vskip -0.2in
\end{figure*}

Language models are typically evaluated on safety benchmarks that include diverse item types probing different aspects of a model’s safety behavior, such as direct harmful requests, adversarial prompts, refusal robustness, multilingual safety questions, and bioweapon-uplift knowledge~\citep{liu2025scalesjustitiacomprehensivesurvey}. As a result, different items provide different amounts of information about a model’s latent safety ability, and some prompts are much more informative than others. At the same time, running full safety evaluations is expensive. A full pass over modern benchmarks can require hundreds of thousands of model responses, yet most of these contribute little to evaluating each model. Only the items where models disagree carry a meaningful ranking signal. For capability benchmarks, fewer than 10\% of items determine most of the leaderboard \citep{rodriguez-etal-2021-evaluation}. Treating items as interchangeable wastes evaluation budget and can obscure meaningful differences between models.

Item Response Theory (IRT) formalizes this asymmetry by modeling responses through item-level parameters (difficulty and discrimination) ~\citep{lalor2019emnlp_pyirt}. Adaptive methods such as Computerized Adaptive Testing (CAT) ~\citep{magis2017cat, cat_intro_meijer_nering_1999} use these parameters to iteratively select the most informative item based on the current ability estimate, thereby allowing fewer evaluations while preserving ranking accuracy and comparable scores.

In this work, we extend the Fluid Benchmarking framework of~\citet{hofmann2025fluid} from capability evaluation to safety evaluation, where adversarially designed prompts, heterogeneous threat models, and skewed response distributions from aligned models pose distinct challenges. Fluid Benchmarking is an IRT-based method that adaptively selects highly informative items tailored to each model under evaluation, reducing cost without sacrificing ranking quality. We apply it to six safety benchmarks spanning direct harmful requests, refusal robustness, adversarial prompting, and bioweapon-uplift settings. We extend the framework with three new IRT-based methods and two adapted non-IRT methods that produce static item subsets reusable across models, without per-model selection.


We show that Item Response Theory recovers interpretable structure in safety benchmarks. Fitted item parameters track empirical response behavior, and latent ability estimates produce stable rankings even where safety scores saturate. We describe how benchmark items with highly variable discrimination power broadly benefit from the IRT-based method.

We demonstrate that adaptive item selection can substantially reduce evaluation cost in safety settings while preserving model rankings, achieving more than an order-of-magnitude reduction in evaluation cost across several benchmarks.

We show that static subset selection enables practical, low-cost safety benchmarking while preserving ranking quality. Compact, model-agnostic subsets accurately reproduce full-benchmark leaderboards while remaining compatible with standard evaluation pipelines.

Figure~\ref{fig:recipe__} provides an overview of our approach and the 
steps taken in this work.

\section{Related Work}
\label{sec:related}

\paragraph{Safety Benchmarks for Language Models.} Recent research has produced numerous safety benchmarks designed to address diverse adversarial scenarios, threat models, and capabilities to evaluate large language models. These efforts span several distinct axes: direct-harm refusal~\citep{mazeika2024harmbench, vidgen2024simplesafetytests, wang2023donotanswer, xie2024sorrybench, ganguli2022redteaming}, over-refusal of benign prompts~\citep{rottger2024xstest}, jailbreak robustness~\citep{souly2024strongreject, zou2023advbench,
chao2024jailbreakbench}, risk-taxonomy coverage~\citep{zeng2024airbench2024safetybenchmark, zhang2024safetybenchevaluatingsafetylarge, wmdp_benchmark}, real-user toxicity~\citep{lin2023toxicchat, ji2024beavertails, han2024wildguard}, and agentic tool use~\citep{zhan2024injecagentbenchmarkingindirectprompt,
andriushchenko2025agentharm}. Aggregator efforts such as HELM Safety~\citep{liang2023helm, helm-safety} and TrustLLM~\citep{sun2024trustllm} consolidate subsets of these into multi-task leaderboards. Despite their diversity, all of these benchmarks share a common evaluation paradigm: items are administered uniformly to all models, and scores are computed as simple averages over binary or thresholded judgments. To our knowledge, no prior work leverages item-level psychometric structure to reduce the cost of safety evaluation.

\paragraph{Efficient Evaluation of LLMs.} Several recent studies contributed methods that reduce the capability-evaluation cost by selecting smaller, more informative item subsets. tinyBenchmarks~\citep{polo2024tinybenchmarks} fits an IRT model to leaderboards such as MMLU~\cite{wang2024mmluprorobustchallengingmultitask} and uses a small anchor set to estimate full-benchmark scores; Anchor Points~\citep{vivek2024anchor} selects medoids of items clustered by class-conditional model agreement; and DISCO~\citep{rubinstein2026discodiversifyingsamplecondensation} selects items maximizing Jensen-Shannon divergence across class-conditional probabilities. Most relevant to us, Fluid Benchmarking~\citep{hofmann2025fluid} integrates item response theory with computerized adaptive testing~\citep{magis2017cat, cat_intro_meijer_nering_1999} to construct capability benchmarks, and also introduces a validity metric grounded in cross-benchmark accuracy. We extend this framework to safety evaluation, which differs in some respects, such as: (a) aligned responses are sparse, and ceiling effects are substantial; (b) cross-benchmark validity is ill-posed because safety benchmarks target heterogeneous harm dimensions rather than a unified latent competence; (c) item discrimination varies widely across benchmarks. We additionally adapt Anchor Points~\citep{vivek2024anchor} and DISCO~\citep{rubinstein2026discodiversifyingsamplecondensation} to the binary-only regime of our safety evaluation, where class-conditional probabilities are unavailable for closed-weight models.

\paragraph{IRT and Adaptive Testing.} IRT models response data through latent ability and item-level parameters (\textit{difficulty, discrimination})~\citep{lord1980applications, embretson2000irt, baker2004item}, and in the context of language models, IRT has been used to study dataset quality, annotator behavior, and the informativeness of benchmarks~\cite{lalor2019emnlp_pyirt, rodriguez-etal-2021-evaluation}. CAT builds on IRT by sequentially selecting items that maximize information about the current ability estimate~\citep{cat_intro_meijer_nering_1999, magis2017cat}. Our work shows that IRT-based adaptive testing remains effective on highly heterogeneous safety benchmarks. The fitted item parameters also serve as diagnostics for benchmark redundancy, ranking resolution, and evaluation cost.

\section{Methodology}
\label{sec:methodology}

In the following subsections, we first introduce the safety benchmarks used in this work and the 2PL IRT model~\citep{baker2004item} we fit to each benchmark (\S~\ref{sec:benchmarks}-\ref{sec:irt_setup}). We then present the adaptive safety testing procedure, based on the method proposed by \citeauthor{hofmann2025fluid} and a static subset selection approach (\S~\ref{sec:adaptivesafetytesting}-\ref{sec:statissubsetselection}). Finally, we describe the \texttt{Agreement} metric, our performance measure used in both approaches (\S~\ref{sec:eval_metrics}). 

\subsection{Safety Benchmarks}
\label{sec:benchmarks}

\begin{table}[ht]
\centering
\caption{Summary of safety benchmarks used in our study. “Models” and “Items” give the number of evaluated models and benchmark items, respectively; “Obs.” is their product, i.e., the total number of model-item observations. See Appendix~\ref{app:dataset} for more details.}
\begin{tabular}{lrrr}
    \toprule
    \textbf{Benchmark} & Models & Items & Obs. \\
    \midrule
    AIR-Bench 2024   & 82 & 5,694  & 467k \\
    Anthropic Red Team& 77 & 1,000  & 77k  \\
    HarmBench          & 77 & 400   & 30.8k \\
    SafetyBench        & 29 & 11,435 & 333k \\
    SimpleSafety       & 77 & 100   & 7.7k \\
    WMDP (Bio)         & 29 & 1,273  & 36.9k \\
    \bottomrule
\end{tabular}
\label{tab:benchmark_stats}
\vskip -0.1in
\end{table}

Our evaluation spans six safety benchmarks that together cover the main response formats used in the field: open-ended generation judged for harmful content, multiple-choice questions probing unsafe knowledge, and graded-severity assessments of policy violations. AIR-Bench 2024~\citep{zeng2024airbench2024safetybenchmark} covers adversarial prompts across 314 fine-grained risk categories with graded severity judgments. Anthropic Red Team~\citep{ganguli2022redteaming} consists of human-generated adversarial dialogues with broad topical coverage. HarmBench~\citep{mazeika2024harmbench} targets direct harmful requests for automated red-teaming and refusal robustness. SafetyBench~\citep{zhang2024safetybenchevaluatingsafetylarge}  is a multilingual multiple-choice benchmark spanning seven safety categories.\footnote{However, for the purpose of this study, we only use the English version.} SimpleSafetyTest~\citep{vidgen2024simplesafetytests} provides a focused suite for rapid screening across five harm areas. WMDP (Bio)~\citep{wmdp_benchmark} is a multiple-choice probe of biology knowledge correlated with bioweapon uplift (here, unsafe behavior corresponds to answering correctly). Table~\ref{tab:benchmark_stats} lists the number of models, items, and the responses collected for each benchmark. We provide more dataset details in Appendix~\ref{app:dataset}.

\subsection{Item Response Theory Setup}
\label{sec:irt_setup}

For each benchmark of Table~\ref{tab:benchmark_stats}, the evaluation output is organized as a binary matrix $\mathbf{X} \in \{0,1\}^{M \times N}$, where $M$ is the number of models and $N$ is the number of benchmark items. Each entry $x_{mj}$ encodes whether model $m$ responded safely ($x_{mj} = 1$) or unsafely ($x_{mj} = 0$) to item $j$. We model the probability that model $m$ responds safely to item $j$ using the Two-Parameter Logistic (2PL) IRT model~\citep{baker2004item}. The 2PL specifies this probability as a function of a unidimensional latent \textbf{safety ability} $\theta_m \in \mathbb{R}$ for each model, and two item parameters: \textbf{discrimination} $a_j\in \mathbb{R}^+$ and \textbf{difficulty} $b_j \in \mathbb{R}$:
\begin{equation}
P(x_{mj} = 1 \mid \theta_m, a_j, b_j) = \sigma\bigl(a_j \cdot (\theta_m - b_j)\bigr)
\label{eq:2pl}
\end{equation}
where $\sigma(z) = (1 + e^{-z})^{-1}$ is the logistic function. Intuitively, $b_j$ captures item difficulty (higher values mean even relatively safe models tend to respond unsafely) and $a_j$ captures how sharply the item separates safer from less safe models. We refer to $\theta_m$ as the \textbf{model's latent safety ability} and to the empirical fraction of items handled safely as the model's safety score. Both are summaries of safety-aligned behavior rather than correctness against a ground truth.

We use the 2PL rather than the 1PL (Rasch) model~\citep{embretson2000irt} as safety items vary substantially in discrimination. The 1PL constrains all items to have equal discrimination, which is inappropriate given the observed variation in how sharply safety items separate models. Fitting the model yields both item parameters $(a_j, b_j)$. 

We provide details on the IRT model fitting procedure in Appendix~\ref{app:irt_estimation}.


\subsection{Adaptive Safety Testing}
\label{sec:adaptivesafetytesting}

\begin{algorithm}
    \small
    \caption{Adaptive Safety Evaluation}
    \label{alg:cat}
    \KwIn{item collection $\mathcal{I}$, model $r_m$,\\
    \hspace{1.5em}selection criteria $\mathcal{S} \in$ 
    $\{\text{Random}, \text{MaxFisherInformation}\}$,\\
    \hspace{1.5em}estimator $\mathcal{E} \in \{\text{IRT}, \text{Average Safety Score}\}$,\\
    \hspace{1.5em}budget $K \le |\mathcal{I}|$}
    \KwOut{score $\hat{\theta}$}
    $\hat{\theta} \gets 0$,\quad $\mathcal{A} \gets \emptyset$\;
    \While{$|\mathcal{A}| < K$}{
        Select $j^{*} \gets \mathcal{S}(\mathcal{I} \setminus \mathcal{A}, \hat{\theta})$\;\\
        Observe $x_{j^{*}} \gets r_m(j^{*})$\;   \\
        $\mathcal{A} \gets \mathcal{A} \cup \{j^{*}\}$\; \\
        Update $\hat{\theta}$ via $\mathcal{E}$ on $\{x_j\}_{j \in \mathcal{A}}$\;
    }
    \Return $\hat{\theta}$

\end{algorithm}

Evaluating model safety over a fixed pool of items is inherently redundant: many items probe similar behaviors and yield overlapping information. As a result, the order in which items are administered can significantly affect the efficiency of evaluation. An adaptive strategy can exploit this structure by prioritizing items that are most informative given the model’s observed behavior, reducing redundancy while focusing evaluation on regions where the model is uncertain or prone to failure. We describe this approach using a computerized adaptive testing framework, as detailed below.

\subsubsection{The CAT Loop}
\label{sec:catloop}
Computerized adaptive testing (CAT) is a sequential procedure that dynamically administers items based on a model's estimate of ability~\citep{cateducational}. Algorithm~\ref{alg:cat} summarizes the loop: at each step, the next item is selected using a selection criterion, the model's response is observed, and the ability estimate $\hat{\theta}$ is updated. \textbf{Fluid Benchmarking} is an algorithm introduced by~\citeauthor{hofmann2025fluid} that applies maximum a posteriori (MAP) estimation on a pre-calibrated IRT model to estimate abilities (see Appendix~\ref{app:map_ability_est} for details). At each step, the item with the maximum Fisher information (MFI) is selected. Under the 2PL IRT model, item $j$'s Fisher information at $\theta$ is
\begin{equation}
I_j(\theta;a_j,b_j) = a_j^{2} P_j(\theta;a_j,b_j)\bigl(1 - P_j(\theta;a_j,b_j)\bigr),
\label{eq:mfi}
\end{equation}
and the MFI rule selects $j^\star = \arg\max_{j \notin \mathcal{A}} I_j(\hat{\theta})$, where $\mathcal{A}$ is the selected item pool and $\hat{\theta}$ is the estimated ability. This rule favors items that are (i) well-matched to the current ability level and (ii) highly discriminating (large $a_j$), thereby maximizing information gain and reducing posterior uncertainty in $\hat{\theta}$.

\subsection{Static Subset Selection}
\label{sec:statissubsetselection}

In the adaptive testing procedure of Algorithm~\ref{alg:cat}, a new subset is built iteratively and tailored to each model under consideration. An alternative approach to efficiently test language models is to select a static, informative subset of items from a given benchmark. In this case, the items are the same for all language models, but they are carefully selected so that the performance on these items is close to that on the full benchmark. The advantage, compared to dynamic selection, is that items are available out of the box, since the subset is fixed \textit{a priori}. The expected trade-off is that a specific model's performance may be less accurate, with a higher likelihood of lower-information items being administered to some models.

In the remainder of this section, we present various procedures for selecting a subset of informative items. Some of our methods rely on IRT fits and ability estimates, while others adapt existing approaches based on more traditional accuracy estimation. Some of the models evaluated on a given benchmark are used as a training sample to tune the approach. The remaining models, the test sample, serve to understand the out-of-sample performance on the subset. For all benchmarks, we discard items with response variance equal to or less than $0.01$\footnote{Calculated as the variance of a binomial distribution, since the scores are binarized.} across models.

We first describe three IRT-based methods. In the following, we indicate by $\theta_m$ the ability of train model $m$ obtained from the MAP method described in Appendix~\ref{app:map_ability_est} using all the benchmark items.

    \textbf{Total Fisher Information.} Each benchmark item $j$ is ranked based on the total Fisher information (Equation~\ref{eq:mfi}) summed across models: $I_j^{Total Fisher}=\sum_m I_j(\theta_m)$. A subset of size $k$ is formed by progressively joining the first $k$ items in decreasing order of total Fisher information. 
    
    \textbf{Marginal Fisher Information.} A more refined approach progressively builds a subset by adding, at each step, the item that most reduces the sum of standard errors of ability estimates across the training models, given the items already added. After $k$ steps, this greedy procedure has selected items minimizing $\sum_m \text{SE}_k(\theta_m) = \sum_m \frac{1}{\sqrt{\sum_{j=1}^{k} I_j(\theta_m)}}$.\footnote{We use the formula of the standard error of the maximum likelihood estimator, which is a common approximation to the standard error of the MAP estimate.}
    
    \textbf{Marginal Fisher Information - $b$ quartile constrained.} This method is nearly identical to Marginal Fisher Information, but it adds constraints on item selection. Items are first stratified into difficulty bins quartiles (using the IRT difficulty parameter $b$). The items are selected in a round-robin fashion across quartiles, by picking items that minimize $\sum_m\text{SE}_k(\theta_m)$. This ensures that the resulting list contains approximately proportional coverage across the full spectrum of difficulty.

In addition to these, we implement simplified versions of the \textbf{Anchor Point}~\cite{vivek2024anchor} and \textbf{DISCO}~\cite{rubinstein2026discodiversifyingsamplecondensation} techniques. Anchor Point selects a subset of items that captures overall diversity based on the train model scores. The DISCO method selects informative items where the model responses are diverse. These methods were developed for general capability benchmarks and rely on access to class-conditional predictive probabilities for each answer choice. This limits their applicability in our setting, where such probabilities (or underlying logits) are unavailable, as is often the case when evaluating closed-weight models. We therefore implement a simple adaptation for our case, described in Appendix~\ref{app:anchor_disco}.

\subsection{Performance Evaluation}
\label{sec:eval_metrics}

For both the dynamic (\S\ref{sec:adaptivesafetytesting}) and static subset selection (\S\ref{sec:statissubsetselection}), we evaluate the model's performance on the selected subset against that on the full subset. For the set of all benchmark items $B$ and a subset $B_k \subseteq B$ of size $k$, we define the \texttt{Agreement} as
{%
\abovedisplayskip=7pt
\belowdisplayskip=7pt
\begin{equation}
\label{eq:agreement_def}
\texttt{Agreement}(P, B, B_k)=\rho_\texttt{Spearman}\bigl(P(B_k),R(B)\bigr),
\end{equation}
}

where $P$ is a performance metric that evaluates the models' performance on the subset, and $R$ is a performance metric of the models on the full set. Spearman’s $\rho$ essentially measures safety ranking fidelity, with values close to 1 indicating that the relative safety across models can be precisely reproduced by evaluating on $k$ items. In the rest of this paper, $P$ is the estimated MAP ability (Appendix~\ref{app:map_ability_est}) for IRT-based methods (Fluid Benchmarking, Total Fisher, Marginal Fisher, and Marginal Fisher - $b$ quartile). $P$ is instead the safety score for the Anchor Point and DISCO methods. $R$ is the safety score for all methods.

We measure \texttt{Agreement} for all selection methods across different subset sizes $k$ on each benchmark. To enable a fair comparison, we randomly split the models for each benchmark into a training set and a held-out test set containing 10 models. The training models are used both to calibrate the IRT parameters and to construct the static subset, while \texttt{Agreement} is evaluated exclusively on the held-out test models. Results for both the dynamic and static selection methods are averaged over 70--100 random train--test splits, depending on the benchmark size.

We additionally compare the subset selection methods introduced in \S\ref{sec:adaptivesafetytesting} and \S\ref{sec:statissubsetselection} against two baselines. In the \textbf{Random} baseline, $k$ items are sampled uniformly at random, and the safety score is computed directly from the responses to those items. In the \textbf{Random IRT} baseline, $k$ items are likewise sampled uniformly at random, but model abilities are estimated via MAP from the responses to those items. The two baselines thus differ in whether models are ranked by raw safety scores or by MAP ability estimates (the $P$ term in Equation~\ref{eq:agreement_def}).

\section{Results}
\label{sec:results}

We organize our results around three questions: (1) Does the IRT model recover meaningful structure from safety benchmarks? (2) Can adaptive item selection exploit that structure to reduce evaluation cost? (3) Can static subsets preserve ranking quality without per-model adaptivity? 


\subsection{IRT Model Fit} 
\label{sec:irt_model_fit}
\begin{figure}[t]
\centering
\includegraphics[width=\linewidth]{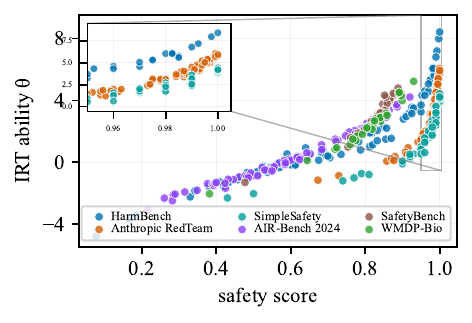}
\caption{The 2PL model recovers rank ordering with Spearman's $\rho \geq 0.97$ on every benchmark (each point represents a model). The inset shows that IRT safe ability continues to separate models where the safety score saturates above $0.95$.}
\label{fig:irt_ability}
\vskip -0.25in
\end{figure}

Figure~\ref{fig:irt_ability} plots the MAP ability estimates $\hat{\theta}_m$ against each model's safety score. The 2PL fit recovers the rank ordering induced by safety score on every benchmark, with Spearman’s $\rho$ in the $[0.97, 1.00]$ range across benchmarks. For each item, the predicted safety scores match the observed scores with a root mean squared error (RMSE) of at most 0.04 across all benchmarks (Appendix~\ref{app:irt_fit}). Beyond rank agreement, IRT ability estimates provide additional resolution when safety scores saturate: on benchmarks where several models score above $0.95$, the fitted $\hat{\theta}_m$ values still distinguish between them, yielding a more fine-grained ranking among the safest models. On Anthropic Red Team, 67 out of 77 models achieve safety scores above $0.95$, yet IRT separates them across a $\theta$ range from $1.19$ to $6.06$. As a result, models that appear similarly safe according to the raw safety score (for example, $0.957$ versus $0.999$) can still differ by nearly five standard units in latent safety ability. HarmBench exhibits an even larger separation: 19 models with safety scores between $0.95$ and $1.00$ are distributed across a broad range of ability values. SimpleSafety shows the same effect on a smaller scale. By contrast, AIR-Bench 2024, SafetyBench, and WMDP-Bio contain no models with safety scores above $0.95$, indicating that these benchmarks are less susceptible to ceiling saturation for the models considered here.

\subsection{Item Discrimination Varies Widely Across Safety Benchmarks}
\label{sec:discrimination_acrros_safety_bench}

\begin{figure}[t]
\centering
\includegraphics[width=\linewidth]{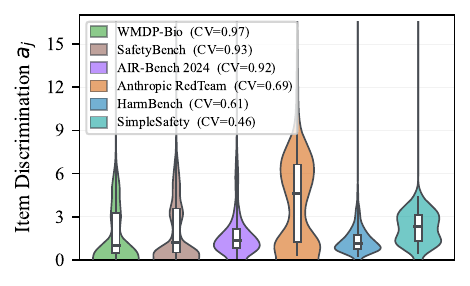}
\caption{Distribution of fitted 2PL item discrimination $a_j$ per benchmark, shown as reflective kernel densities (clipped at zero, with equal maximum widths). Inner boxes mark the interquartile range and median; whiskers span the 1st--99th percentiles. Benchmarks are ordered by coefficient of variation (\texttt{CV}), from $0.46$ (SimpleSafety) to $0.97$ (WMDP-Bio). High \texttt{CV} indicates that a few items carry most of the discriminative signal, while low \texttt{CV} indicates that items contribute roughly uniformly.}
\label{fig:discrimination_dist}
\vskip -0.2in
\end{figure}

A central question for efficient evaluation is whether safety benchmarks contain items with sufficiently diverse discriminative power to support adaptive selection. Figure~\ref{fig:discrimination_dist} presents the distribution of fitted 2PL discrimination parameters $a_i$, summarized using the coefficient of variation (\texttt{CV}), defined as the ratio of the standard deviation to the mean. This normalization accounts for differences in average discrimination across benchmarks.

The benchmarks fall into three broad regimes. (1) \textit{High variance with long right tails:} WMDP-Bio (\texttt{CV}~$=0.97$), SafetyBench (\texttt{CV}~$=0.93$), and AIR-Bench 2024 (\texttt{CV}~$=0.92$) exhibit low median discrimination together with a sparse tail of highly discriminative items, precisely the setting in which adaptive selection has the greatest opportunity to improve efficiency. (2) \textit{Moderate variance:} Anthropic Red Team (\texttt{CV}~$=0.69$) and HarmBench (\texttt{CV}~$=0.61$) display an intermediate spread without a pronounced extreme tail, suggesting more modest efficiency gains from adaptive selection. (3) \textit{Low variance:} In SimpleSafety (\texttt{CV}~$=0.46$), items provide relatively uniform information, limiting the potential advantage of MFI selection. Accordingly, benchmarks with larger \texttt{CV} are expected to exhibit greater gains from MFI-based adaptive selection, whereas benchmarks with more homogeneous discrimination should show similar performance for Random IRT and Fluid Benchmarking.

\subsection{Adaptive and Static Item Selection}
\label{sec:item_selection}
\begin{figure}[t]
    \centering
    \centering
    \includegraphics[width=\linewidth]{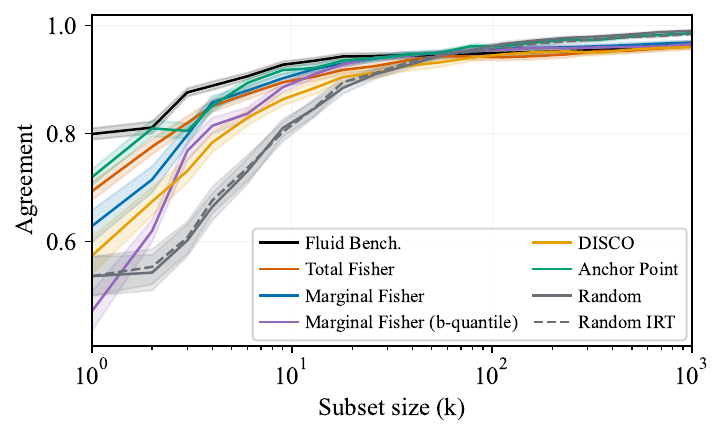}

    \caption{\texttt{Agreement} as a function of subset size for various subset selection methods for the AIR-Bench 2024 benchmark. We report the mean \texttt{Agreement} across different train-test split runs. The shaded regions denote the 1 standard deviation intervals of the mean across runs.}
    \label{fig:agreement_airbench}
    \vskip -0.2in
\end{figure}

In Figure~\ref{fig:agreement_airbench} we report the \texttt{Agreement} metric for the Fluid Benchmarking method and the various static selection methods described in \S\ref{sec:statissubsetselection} for the AIR-Bench 2024 benchmark and for various subset sizes $k$. Appendix~\ref{app:appendix_agreement} reports the same plots for all the benchmarks. Figure~\ref{fig:agreement_airbench} shows that the adaptive procedure outperforms all other selection methods at low $k$, consistent with its more targeted and model-specific design. For AIR-Bench 2024, the modified Anchor Point, Total Fisher, and Marginal Fisher approaches perform similarly, achieving high correlation with the safety-score-induced ranking using only a small number of items. Notably, all methods outperform random selection at small values of $k$. However, the \texttt{Agreement} of a random subset converges toward that of the other selection methods at approximately $k=50$. We interpret this as follows: IRT-based approaches prioritize high-discrimination items (since Fisher information scales with the square of item discrimination), but AIR-Bench 2024 contains only a limited number of such items in its tail. As a result, IRT-based methods quickly exhaust the pool of high-discrimination items. At larger $k$, however, random sampling becomes increasingly likely to include those same high-discrimination items by chance, thereby diminishing the advantage of deliberate selection. This convergence, therefore, reflects a property of the dataset, the scarcity of highly discriminating items, rather than a limitation of IRT-based methods themselves. We further highlight that Random and Random IRT are nearly indistinguishable, suggesting that IRT ability estimation and raw safety score yield consistent rankings when items are sampled randomly. This supports the finding in Section~\ref{sec:irt_model_fit} that the IRT formulation is a coherent extension of the safety score. The high \texttt{Agreement} achieved even at small $k$ further highlights the substantial redundancy among dataset items, as a small random subset is already sufficient to approximate the full-set evaluation.

\subsubsection{Efficiency Gains Across Benchmarks}
\label{sec:efficiency}

\begin{table*}[ht]
\centering
\caption{\texttt{Agreement} between subset and full-benchmark scores across benchmarks and selection methods. For each benchmark, the left $k$ is the first value at which Fluid Benchmarking reaches $90\%$ \texttt{Agreement} (a conservative estimate over a logarithmic scan); the right $k$ is the last value before the best random baseline (Random or Random IRT) matches Fluid Benchmarking's \texttt{Agreement} (see \S\ref{sec:efficiency}).}
\label{tab:efficiency}
\small
\setlength{\tabcolsep}{4pt}
\begin{tabular}{lcccccccccccc}
\toprule
& \multicolumn{2}{c}{\textbf{AIR-Bench 2024}} & \multicolumn{2}{c}{\textbf{Ant.\ Red Team}} & \multicolumn{2}{c}{\textbf{HarmBench}} & \multicolumn{2}{c}{\textbf{SafetyBench}} & \multicolumn{2}{c}{\textbf{SimpleSafety}} & \multicolumn{2}{c}{\textbf{WMDP}} \\
\cmidrule(lr){2-3}\cmidrule(lr){4-5}\cmidrule(lr){6-7}\cmidrule(lr){8-9}\cmidrule(lr){10-11}\cmidrule(lr){12-13}
$N$ & \multicolumn{2}{c}{5,694} & \multicolumn{2}{c}{1,000} & \multicolumn{2}{c}{400} & \multicolumn{2}{c}{11,435} & \multicolumn{2}{c}{100} & \multicolumn{2}{c}{1,273} \\
\cmidrule(lr){2-3}\cmidrule(lr){4-5}\cmidrule(lr){6-7}\cmidrule(lr){8-9}\cmidrule(lr){10-11}\cmidrule(lr){12-13}
$k$ & 6$^\dagger$ & 38$^\ddagger$  & 51$^\dagger$ & 260$^\ddagger$  & 17$^\dagger$ & 9$^\ddagger$  & $>$3,000$^*$ & 68$^\ddagger$  & 20$^\dagger$ & 63$^\ddagger$  & 78$^\dagger$ & 234$^\ddagger$  \\
$k/N$ & 0.1\% & 0.7\% & 5\% & 26\% & 4\% & 2\% & $>$26\% & 0.6\% & 20\% & 63\% & 6\% & 18\% \\

\midrule
Fluid Benchmarking         & 0.91 & 0.94 & 0.91 & 0.97 & 0.90 & 0.88 & 0.82 & 0.70 & 0.91 & 0.98 & 0.90 & 0.92 \\
Marginal Fisher            & 0.88 & 0.95 & 0.87 & 0.97 & 0.90 & 0.88 & 0.82 & 0.68 & 0.90 & 0.98 & 0.85 & 0.92 \\
Marginal Fisher ($b$-quant.) & 0.84 & 0.94 & 0.89 & 0.96 & 0.90 & 0.87 & 0.91 & 0.84 & 0.84 & 0.97 & 0.87 & 0.91 \\
Total Fisher               & 0.87 & 0.94 & 0.68 & 0.87 & 0.83 & 0.81 & 0.81 & 0.64 & 0.83 & 0.96 & 0.85 & 0.92 \\
Anchor Point               & 0.89 & 0.95 & 0.77 & 0.98 & 0.94 & 0.90 & 0.92 & 0.79 & 0.87 & 0.99 & 0.86 & 0.89 \\
DISCO                      & 0.83 & 0.93 & 0.91 & 0.98 & 0.90 & 0.87 & 0.91 & 0.64 & 0.92 & 1.00 & 0.81 & 0.90 \\
Random                     & 0.73 & 0.93 & 0.78 & 0.96 & 0.92 & 0.88 & 0.97 & 0.70 & 0.75 & 0.96 & 0.86 & 0.91 \\
Random IRT             & 0.74 & 0.93 & 0.79 & 0.91 & 0.91 & 0.87 & 0.97 & 0.71 & 0.75 & 0.94 & 0.80 & 0.88 \\
\bottomrule
\end{tabular}
\\[4pt]
\raggedright \footnotesize$N$ = full benchmark size;\quad $k$ = subset size.\\
\raggedright\footnotesize $^\dagger$ Minimum $k$ for Fluid Benchmarking $\geq 90\%$ \texttt{Agreement};\quad 
$^\ddagger$ last $k$ before best random baseline matches Fluid Benchmarking. \\
\raggedright\footnotesize $^*$ At this value, we stop scanning the $k$ values since random methods are already outperforming all other methods.
\end{table*}


For each benchmark, Table~\ref{tab:efficiency} reports \texttt{Agreement} at two values of $k$: the smallest $k$ at which Fluid Benchmarking first reaches $90\%$ \texttt{Agreement} (left), and the largest $k$ before the best random baseline (Random or Random IRT) attains the same \texttt{Agreement} (right). Since we scan $k$ logarithmically for computational efficiency, both choices are conservative since the true crossover points may occur at smaller $k$ than reported. The former measures how efficiently IRT-based methods compress a benchmark; the latter quantifies how much more data random selection requires to reach equivalent quality. Fluid Benchmarking is the most consistent method across benchmarks, reaching $90\%$ \texttt{Agreement} at remarkably small subset sizes (as low as $0.1\%$ of the full benchmark on AIR-Bench 2024). Among the other IRT-based methods, Marginal Fisher is the closest competitor, matching or even slightly exceeding Fluid Benchmarking once $k$ grows beyond the $90\%$ threshold (e.g., $0.95$ vs.\ $0.94$ on AIR-Bench 2024 at $k=38$), though it lags at the smaller $k$ anchors. Total Fisher is the weakest IRT variant, dropping noticeably below Fluid Benchmarking on several benchmarks (e.g., $0.68$ vs.\ $0.91$ on Anthropic Red Team at $k=51$). Non-IRT methods (Anchor Point, DISCO) and random baselines are markedly less reliable, often requiring several times more items to match Fluid Benchmarking's \texttt{Agreement}. Two benchmarks deviate from this pattern: on HarmBench, even random selection matches IRT-based methods, suggesting low inter-model variance makes the benchmark inherently easy to compress; on SafetyBench, random selections reach $90\%$ \texttt{Agreement} at $k\approx556$, faster than any other methods, though IRT-based methods still substantially outperform random selection at small $k$ (e.g., \texttt{Agreement} $\approx 0.71$ vs.\ $\approx 0.46$ at $k=19$, see Appendix~\ref{app:appendix_agreement}). Assuming evaluation cost scales linearly with the number of items, these compression rates translate directly into substantial cost savings. On AIR-Bench 2024, adaptive selection achieves $99.9\%$ savings and static selection $99.8\%$ for a $90\%$ \texttt{Agreement}. Across the remaining benchmarks where the $90\%$ \texttt{Agreement} threshold is reached, savings range from $80\%$ to $95\%$ for Fluid Benchmarking (adaptive) and from $80\%$ to $92\%$ for Marginal Fisher (static). Both figures exclude HarmBench, where IRT-based methods perform comparably to random selection.


\section{Discussion}
\label{sec:discussion}

Taken together, our results indicate that IRT transfers well from model evaluation to the safety context. Notably, the 2PL parameters are interpretable: items with higher difficulty are answered safely by a smaller fraction of models, and the IRT calibration reproduces benchmark scores with high fidelity. Hence, the latent-variable assumptions underlying IRT appear reasonable, at least for the one-shot benchmarks considered here.

\textbf{\texttt{CV} is a rule-of-thumb diagnostic.} The coefficient of variation (\texttt{CV}) of item discrimination can serve as a lightweight diagnostic for predicting when adaptive item selection is likely to outperform random baselines. Fluid Benchmarking~\cite{hofmann2025fluid} assumes adaptive selection is generally beneficial but does not characterize when or by how much. Our \texttt{CV} analysis addresses this gap: benchmarks with higher \texttt{CV} tend to have a sparse tail of highly discriminative items that adaptive methods can exploit, while low \texttt{CV} suggests more uniform item informativeness and thus smaller gains over random selection. Our results are broadly consistent with this interpretation. On high-\texttt{CV} benchmarks (AIR-Bench 2024, WMDP, SafetyBench, and Anthropic Red Team), IRT-based methods consistently outperform random baselines at small $k$. However, \texttt{CV} does not fully explain all variation: SimpleSafety has a low \texttt{CV} yet IRT methods retain a clear advantage over random selection, while HarmBench has a moderate \texttt{CV} yet random selection matches IRT methods. This suggests that \texttt{CV} captures an important but incomplete picture of benchmark compressibility, and that other factors, such as the overall level of inter-model agreement or benchmark size, may also play a role.


Beyond aggregate statistics, item-level IRT parameters carry diagnostic value relative to the set of models under evaluation. Difficulty parameters identify items where most models fail to respond safely, while discrimination parameters identify items that best separate models within that same set. However, these parameters are specific to the evaluated model population and should not be interpreted as intrinsic properties of the items themselves. With this caveat in mind, IRT parameters could nonetheless support a practical workflow for benchmark developers seeking to understand which items drive differences among a specific cohort of models, or to construct compact variants tailored to that cohort.

\textbf{IRT approaches are best for repeated, costly evaluations.} IRT-based evaluation is most useful when a benchmark will be reused across many models over time, since the upfront cost of fitting item parameters is amortized over repeated evaluations, and each new model can be assessed with fewer items. It is also well-suited to settings where evaluation is expensive and cost matters, such as benchmarks requiring human judges or API calls. In such settings, even moderate reductions in item count can expand what becomes feasible in practice. On the other hand, for one-off evaluations or benchmarks with few items, the overhead of calibrating an IRT model may not be justified, and running the full benchmark will avoid introducing modeling assumptions. 


\textbf{Adaptive selection maximizes accuracy; static subsets favor simplicity.} Adaptive item selection yields more robust \texttt{Agreement} with full-benchmark rankings but produces model-specific subsets that cannot be reused across evaluations. Static subset selection trades some accuracy for simplicity: once a subset is identified, it can be applied to any model without additional overhead, making it better suited for continuous or repeated evaluations across evolving model families. Both approaches reduce intra-benchmark redundancy, freeing evaluation budget that can be reallocated to broader benchmark coverage. This is especially valuable in safety contexts, where comprehensive evaluation typically requires running multiple benchmarks. We recommend adaptive selection when evaluation efficiency and ranking fidelity are the primary concern, and static subsets when comparability across models and operational simplicity are preferred.

\textbf{IRT-based methods are more reliable than other static selection methods at low $k$.} Among static selection methods, IRT-based approaches, such as Marginal Fisher, generally achieve higher \texttt{Agreement} with full-benchmark rankings at small subset sizes than non-IRT alternatives such as Anchor Point and DISCO. The gap narrows at larger $k$ where non-IRT methods benefit from increased coverage. Random baselines lag furthest behind, typically requiring more items to reach equivalent \texttt{Agreement}. As safety benchmarks grow in sophistication, testing a wider range of traits with increasingly challenging items, our results suggest that IRT-based methods offer the greatest potential for cost reduction, since their advantage over random baselines is most pronounced precisely in the high-variance, high-difficulty regimes that characterize more demanding benchmarks.

\textbf{Limitations.} Several modeling constraints should be mentioned. First, safety might not constitute a single latent trait: for instance, a model may be robust to direct harmful requests yet vulnerable to prompt injection, and performance often varies across categories such as violence, self-harm, and deception. This motivates the incorporation of multidimensional IRT frameworks in the future. Second, our pipeline binarizes ordinal safety judgments to simplify the fitting process, discarding information about severity and uncertainty. Extending to graded-response or continuous IRT models would yield a more faithful representation of underlying judgments and may improve calibration and ranking performance. Lastly, the study is constrained by scope. In the future, we hope to test a more comprehensive suite of benchmarks and models and explore agentic rather than one-shot settings. We therefore view these results not as a definitive assessment of safety benchmarking, but as evidence that psychometric methods constitute a viable and useful direction for making safety evaluation cheaper and more targeted.

\section{Conclusion}
\label{sec:conclusion}
We adapt Item Response Theory and Computerized Adaptive Testing from capability evaluation to safety benchmarking. Across six safety benchmarks spanning direct harmful requests, multiple-choice, and bioweapon-uplift settings, the IRT model recovers item-level structure that aligns with empirical response behavior and admits useful interpretation. On benchmarks with high variance in item discrimination, adaptive item 
selection reduces the number of administered items by roughly two orders of magnitude. Benchmarks with more uniform item discrimination yield smaller gains, with the magnitude of efficiency improvement itself reflecting structural properties of the benchmark. Static subset selection methods, both IRT-based and score-based, complement the per-model adaptive procedure with model-agnostic alternatives suitable for shareable leaderboards. Overall, our results show that psychometric methods are a viable and useful path toward more frequent, lower-cost safety evaluation throughout the model development lifecycle.



\section*{Acknowledgements}

This research was conducted as part of the Supervised Program for Alignment Research (SPAR) Spring 2026, an AI safety research program.

\bibliography{references}
\bibliographystyle{icml2026}

\appendix
\onecolumn
\section{Dataset Details}
\label{app:dataset}

The dataset model scores from AIR-Bench 2024, Anthropic Red Team, HarmBench, and SimpleSafetyTest are taken from HELM Safety~\cite{helm-safety}. We ran our own evaluation on SafetyBench and WMDP, scoring each model on the multiple-choice items via a standard zero-shot evaluation script. Our experiments were run via the OpenRouter API, costing approximately \$100 to evaluate 29 models over the course of one day.

We binarize model responses to each benchmark item into a common $\{0, 1\}$ coding, where $1$ denotes a safe response and $0$ an unsafe one. For benchmarks from HELM Safety~\cite{helm-safety} that rely on an 
LLM judge to assess response safety (HarmBench, Anthropic Red Team, SimpleSafetyTest, AIR-Bench 2024), we label a response as $1$ if the judge assigns it the highest safety category and $0$ otherwise. For 
multiple-choice benchmarks (WMDP, SafetyBench), we code the safe-aligned response as $1$.

Binarization discards information about response severity (graded judgments collapsed to safe/unsafe) and about the
wrong-answer distribution in multiple-choice settings (only the safe-aligned option is distinguished from all alternatives). We adopt this for two reasons: (a) it ensures consistent treatment across heterogeneous benchmarks, several of which are natively binary; (b) the 2PL IRT model is well-studied and computationally tractable, whereas graded-response variants would require benchmark-specific modeling.

For simplicity in evaluation, we drop multilingual prompts and evaluate only English prompts.

All the IRT fits and subset selection procedures were run on standard commercial CPUs, with the full analysis across all benchmarks and methods completing in approximately 1-2 hours.

\section{IRT Fit Procedure}
\label{app:irt_estimation}

We follow the fitting procedure of \citet{hofmann2025fluid} with minimal modification. The 2PL model (Equation~\ref{eq:2pl}) is fit to each benchmark's full binary response matrix via stochastic variational inference (SVI; \citealt{hoffman2013stochasticvariationalinference}) using \texttt{py-irt}~$\geq$~0.7.0 \citep{lalor2019pyirt} under Python~\texttt{3.11.15}. We use the \texttt{2PL} model variant, trained for 1.000 epochs with the Adam optimizer (learning rate 0.1, the \texttt{py-irt} default). Priors follow the \texttt{py-irt} defaults: discrimination parameters $a_j$ are assigned a $\text{Log-}\mathcal{N}(0,1)$ prior (enforcing positivity), and difficulty parameters~$b_j$ a $\mathcal{N}(0,1)$ prior. Ability estimates~$\hat{\theta}_m$ used in downstream CAT and static subset selection are obtained via Maximum a Posteriori (MAP)~\cite{map_estimation} estimation with a $\mathcal{N}(0,1)$ prior, as described in the following Appendix~\ref{app:map_ability_est}.

\section{Ability Estimation Method: Maximum a Posteriori (MAP) Estimation}
\label{app:map_ability_est}

MAP estimation during CAT finds the posterior mode via damped Newton-Raphson after each new response. Given $n$ administered items with responses $\mathbf{x} = (x_1, \ldots, x_n)$, the score function (first derivative) and observed information (negative second derivative) are:

$$s(\theta) = \frac{\mu_0 - \theta}{\sigma_0^2} + \sum_{j=1}^n a_j (x_j - P_j(\theta))$$\ \text{and}

$$\mathcal{H}(\theta) = \frac{1}{\sigma_0^2} + \sum_{j=1}^n a_j^2 \, P_j(\theta)(1 - P_j(\theta)),$$

where $\mu_0 = 0$ and $\sigma_0 = 1$ are the prior hyperparameters. The Newton-Raphson update is:

$$\theta^{(t+1)} = \theta^{(t)} + \frac{s(\theta^{(t)})}{\mathcal{H}(\theta^{(t)})}$$

with individual step sizes clamped to $|\Delta\theta| \leq 1.0$ logits per iteration to prevent divergence from extreme response patterns. If Newton-Raphson fails to converge within 100 iterations (i.e., $|s(\hat{\theta})| > 10^{-6}$), a bisection fallback is invoked on $[-8, 8]$ for up to 200 iterations. The standard error is:

$$\text{SE}_{\text{MAP}}(\hat{\theta}) = \frac{1}{\sqrt{\mathcal{H}(\hat{\theta})}} = \frac{1}{\sqrt{\frac{1}{\sigma_0^2} + \sum_j a_j^2 P_j(\hat{\theta})(1 - P_j(\hat{\theta}))}}.$$

MAP estimation produces \textbf{posterior modes} and is numerically fast ($O(\text{iterations} \times n)$ per update). The $\mathcal{N}(0,1)$ prior provides beneficial shrinkage early in the test when few items have been administered.

\section{Anchor Point and DISCO Adaptations}
\label{app:anchor_disco}

The Anchor Point~\cite{vivek2024anchor} and DISCO~\cite{rubinstein2026discodiversifyingsamplecondensation} methods are recent approaches for selecting informative subsets of benchmark items. Developed for large capability benchmarks such as MMLU~\citep{wang2024mmluprorobustchallengingmultitask}, both methods rely heavily on class-conditional predictive probabilities. As a result, they are directly applicable only when (i) such probabilities are accessible and (ii) model outputs can be readily mapped to discrete classes. This poses significant limitations in the safety evaluation setting, where it is often desirable to assess closed-weight models (which do not expose internal probabilities) and where safety-relevant outputs may not decompose cleanly into predefined classes without additional classification of the raw model outputs. We have therefore adapted and simplified those methods to be applicable for binary output:

    \textbf{Anchor Points}: The objective is to select a subset of items that best represent the diversity of all items, based on how the train models score them. We therefore built a distance matrix $d_{ij}=1-corr(s_i,s_j)$ where $s_i$ and $s_j$ represent the vectors of scores relative to items $i$ and $j$ respectively for all the models, and $corr$ indicates Pearson correlation. The matrix $d_{ij}$ encapsulates the relative similarity across models for each item pair. A K-medoids (PAM) algorithm is then applied to this distance matrix to select $k$ representative items (medoids) that minimize the total distance between each item and its nearest selected representative. We note that, unlike other methods, the Anchor Points selection metric is not monotonic with $k$, in the sense that items for a subset of size $q<k$ do not necessarily form a subset of the items for a reduced benchmark of size $k$. Model accuracy on a subset is computed as a weighted average of item-level binary scores, where each weight corresponds to the number of items represented by the corresponding medoid.

    \textbf{DISCO}: This method does not rely on item clustering, but rather regards more informative items for which the model responses are diverse. The original method in~\cite{rubinstein2026discodiversifyingsamplecondensation} is not directly usable with the 0 and 1 evaluation scores. We therefore defined a metric similar to the Jensen-Shannon Divergence metric used in the original paper. Our approach uses entropy for a binary outcome as a criterion to identify items with higher model disagreement. The items are therefore ordered based on $H_j(p_j)=-p_j\log p_j-(1-p_j)\log(1-p_j)$, where $p_j$ is the estimated probability of models giving a safe answer to item $j$, estimated as the fraction of models that give a safe answer to $j$. Items with higher entropy are deemed more informative. The subset is built by progressively adding items in decreasing entropy order until $k$ items are added.

\section{IRT Fit Details}
\label{app:irt_fit}

\begin{figure}[ht]
  \centering
  \includegraphics[width=0.9\textwidth]{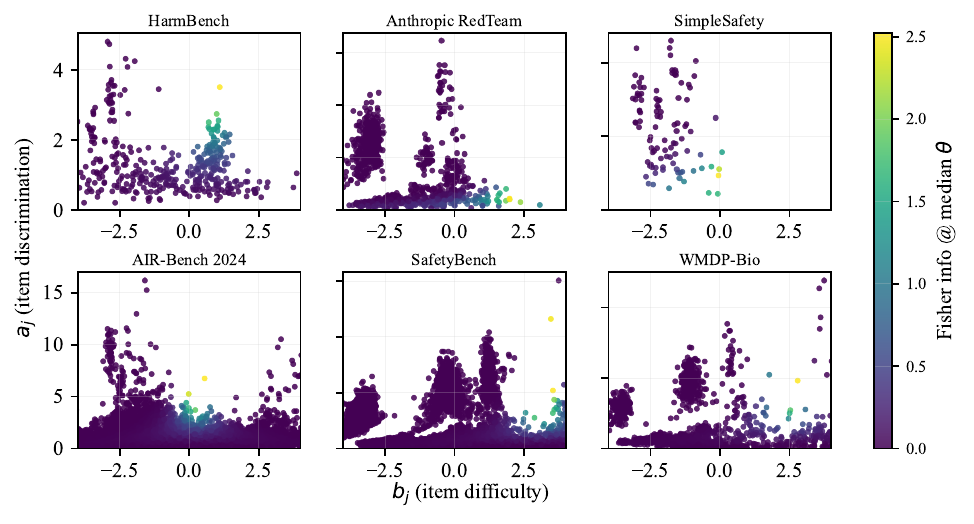}
  \caption{Joint distribution of fitted 2PL item parameters $(a_j, b_j)$ per benchmark, colored by Fisher information $I_j(\tilde{\theta})$ at the median model ability $\tilde{\theta}$. High-information items 
(bright) cluster where $a_j$ is large and $b_j \approx \tilde{\theta}$, while the dense dark region visible on high-\texttt{CV} benchmarks represents redundant items that adaptive selection avoids.}
  \label{fig:discrimination_structure}
\end{figure}

\paragraph{Item-level information structure.} Figure~\ref{fig:discrimination_structure} shows the joint distribution of fitted item parameters $(b_j, a_j)$ per benchmark, with each item colored by its Fisher information $I_i(\tilde{\theta})$ evaluated at the median model ability~$\tilde{\theta}$. The most informative items concentrate in a narrow region: high discrimination and difficulty near~$\tilde{\theta}$, consistent with the 2PL information function $I_j(\theta) = a_j^2\, P_j(\theta)\bigl(1 - P_j(\theta)\bigr)$, which peaks when $\theta \approx b_j$ and scales with~$a_j^2$. On benchmarks with high \texttt{CV} (AIR-Bench 2024, SafetyBench, WMDP-Bio), the vast majority of items sit in a dense low-information floor while a sparse tail of high-$a_j$ items dominates the Fisher budget, the structure that MFI exploits. On SimpleSafety (low \texttt{CV}), Fisher information is distributed more uniformly, explaining the limited advantage of adaptive selection on that benchmark.

\begin{figure}[ht]
    \centering
    \includegraphics[width=0.8\textwidth]{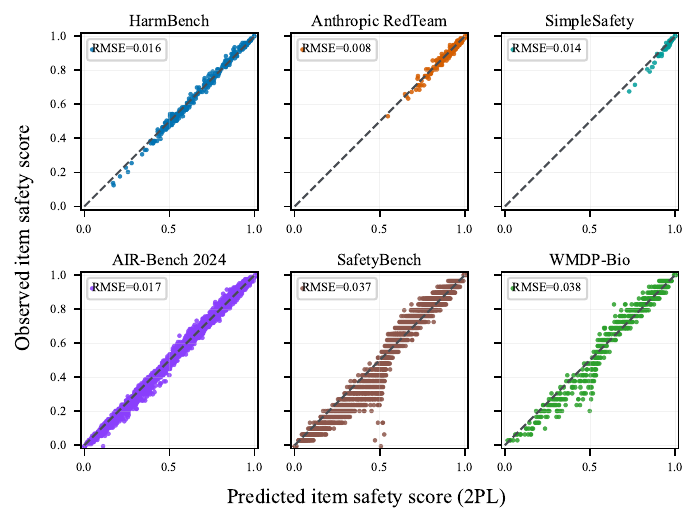}
    \caption{Item-level fit verification of the IRT model. For each item $j$, we use the fitted parameters $(a_j, b_j)$ and abilities $\theta_m$ to compute the predicted probability that each model $m$ responds safely, using Equation~\ref{eq:2pl}. The model-implied item safety score, obtained by averaging these 
probabilities across models, is plotted against the observed safety score on that item.}
    \label{fig:item_level_fit_recovery}
\end{figure}

\paragraph{Fit recovery of marginal item rates.} 
Figure~\ref{fig:item_level_fit_recovery} shows that all six benchmarks 
fall close to the identity line, with RMSE ranging from $0.008$ (Anthropic 
Red Team) to $0.038$ (WMDP-Bio). The 2PL model thus captures how often 
each item is answered safely across the model pool. The concentration 
near the upper-right corner on Anthropic Red Team and SimpleSafety 
reflects the saturation observed in \S\ref{sec:irt_model_fit}, where 
most items live in the high-safety regime. The horizontal banding 
visible on SafetyBench and \textit{WMDP-Bio} is a discretization 
artifact: with $M=29$ models, observed rates can only take values in 
$\{k/M\}_{k=0}^{M}$, which also explains the slightly higher RMSE on 
these two benchmarks. We emphasize that this is a fit-recovery check 
rather than an out-of-sample calibration test; the in-sample item-level 
prediction here supports the model-level results in 
\S\ref{sec:irt_model_fit}.


\section{\texttt{Agreement} Curves Across Benchmarks}
\label{app:appendix_agreement}
\begin{figure}[ht]
    \centering
    \includegraphics[width=0.8\textwidth]{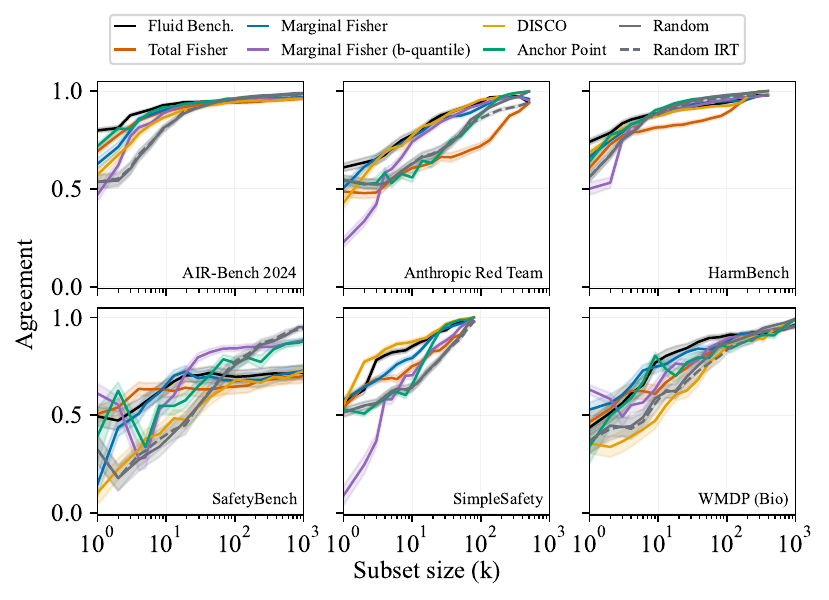}
    \caption{\texttt{Agreement} between subset and full-benchmark performance as a function of the number of administered items, across all six benchmarks. For a full description of the methodology, see \S\ref{sec:eval_metrics}.}
    \label{fig:agreement-all}
\end{figure}

Figure~\ref{fig:agreement-all} reports \texttt{Agreement} curves for all six benchmarks considered in this work. Across benchmarks, we observe a consistent qualitative trend: adaptive item selection via Fluid Benchmarking reaches high agreement with the full-benchmark ranking substantially faster than random baselines. In particular, the adaptive procedure typically achieves strong correlation in the low-sample regime, indicating that a relatively small number of informative items suffices to recover most of the ranking signal. See \S\ref{sec:item_selection} for more details.

\end{document}